\documentclass[a4paper,11pt]{amsart}
\usepackage{multicol,ifthen}
\usepackage{amssymb,stmaryrd}
\usepackage{harvard}
\usepackage{graphicx}

\newcommand{\E}{\mathbb{E}}

\newcommand{\R}{\mathbb R}

\usepackage[colorlinks=false, linktocpage=true,backref,pagebackref,hidelinks]{hyperref}

\newcommand{\beq}{\begin{equation}}
\newcommand{\eeq}{\end{equation}}
\newcommand{\beqarr}{\begin{eqnarray}}
\newcommand{\eeqarr}{\end{eqnarray}}
\newcommand{\beqa}{\begin{eqnarray*}}
\newcommand{\eeqa}{\end{eqnarray*}}
\begin{document}
\renewcommand{\thefootnote}{\fnsymbol{footnote}}
\title[Remarks Weyl -- Cartan, 1920s]{H. Weyl's and E. Cartan's proposals for infinitesimal geometry in the early 1920s}
\author[E. Scholz]{Erhard Scholz$\,^{\dag}$}\footnotetext[2]{\, scholz@math.uni-wuppertal.de\\
\hspace*{7mm}University Wuppertal, Department C, Mathematics and Natural Sciences,\\
\hspace*{6.5mm} and Interdisciplinary Center for Science and Technology Studies} 
\renewcommand{\thefootnote}{\arabic{footnote}}

\date{29. 01. 2010  }

\begin{abstract}
In the early phase of general relativity Elie Cartan and Hermann Weyl thought about the question of  how the role of transformation groups could be transferred from classical geometry (Erlangen program) to differential geometry. They had different starting points and used different techniques, but both generalized the concept of connection arising from Levi-Civita's interpretation of the classical Christoffel symbols as parallel transfer in curved spaces.  Their focus differed and Cartan headed toward a much more general framwork than Weyl (non-holonomous spaces versus scale gauge geometry).  But there also was  an overlap of topics  (space problem) and, at the turn to the 1930s, they arrived at an agreement on how to deal with Cartan's infinitesimal geometric structures. 
\end{abstract}
\thispagestyle{empty}
\maketitle

\section{Introduction}
 Einstein's theory of {general relativity} triggered a multiplicity  of new ideas  in differential geometry. In 1917  Levi-Civita discovered that Einstein's interpretation of the Christoffel symbols in Riemannian geometry  as components of the   gravitational field could be given a geometrical meaning by  the concept of  parallel discplacement. That was the starting point for  investigating of  a whole range of generalized differential geometric structures.   J.A. Schouten and his student D. Struik   studied symbolic methods for establishing an ``absolute calculus'' at Amsterdam. At Z\"urich H. Weyl formed the generalized concept of affine connection, no longer necessarily derived from a Riemannian metric, and generalized the concept of metrical structure by the idea of a gauge metric and a non-integrable scale connection.  A.  Eddington investigated affine and linear connections at Cambridge. At Paris E. Cartan started his  program of bringing Klein's view of geometry  to bear upon differential geometry, and at Princeton the group around   O. Veblen, L.P Eisenhart  and T.Y. Thomas looked for projective structures in differential geometry. Most  of these  geometrical research programs were closely related to attempts of creating a unified field theory of matter, interactions and geometry.\footnote{Accordingly much of the historical literature is  directed at the unified field theory side of the story \cite{Vizgin:UFT,Goenner:UFT,Goldstein/Ritter:UFT}, others look at the geometrical side \cite{Reich:Connection,Gray:Universe,Bourgignon:connex,Scholz:connections,Chorlay:Diss}.}

The upsurge of  new ideas made the 1920/30s a happy time for differential geometry. In this contribution we look at the proposals by {H. Weyl} and {E. Cartan} in the early 1920s. The question of how the Kleinian view of transformation groups could be imported to the differential geometric setting  played a crucial role for both of them. They gave  different answers, although with a certain overlap. Only after further steps of generalizations their views could be subsumed to an even wider frame, that of connections in prinicipal fibre bundles. This was an achievement of the second half of the century, with C. Ehresmann as one of the principal players. It will not be discussed here; here we concentrate on Weyl's and Cartan's respective views in the 1920s.

\section{ Weyl }
\subsection*{Weyl's papers of 1918 and STM} 
In April 1918 A. Einstein presented  Weyl's paper   
{\em Gravitation and electricity} (Gravitation und Elektizit\"at)  \cite{Weyl:GuE},
 to the Berlin Academy of Sciences. He added a short critical comment explaining why he  doubted the reliability of the physical interpretation Weyl gave. The paper contained a scale gauge generalization of Riemannian geometry, with a {\em length connection} expressed by a differential form $\varphi = \sum_i \varphi_i dx^i$ as a crucial ingredient.  Weyl wanted to  identify  the scale connection with the potential of the  electromagnetic field and built  the first geometrically unified theory (UFT) of gravity and electromagnetism on this idea \cite{Vizgin:UFT,ORaifeartaigh:Dawning}. The unification built crucially on the property of $\varphi$ being a  {\em gauge field}.  This idea turned out to be of long lasting importance, although not in its original form.   A few weeks later, a second paper of Weyl followed  in {\em Mathematische Zeitschrift} \cite{Weyl:InfGeo}. It presented the same topic to a mathematical audience and put the Weylian metric into the perspective of a broader view of differential geometry. Here Weyl generalized  Levi-Civita's  idea of parallel displacement in a Riemannian manifold to  that of an {\em affine connection} $\Gamma = (\Gamma^i_{jk})$ (logically) independent of any metric.

The manuscript of Weyl's first book on mathematical physics, 
 {\em Space - Time - Matter} (STM) (Raum -- Zeit -- Materie), 
delivered to the publishing house (Springer) at easter 1918, 
did   {\em not} contain   Weyl's new  geometry and proposal for a UFT. It was prepared from the lecture notes of 
a  course given in summer semester 1917 at the Polytechnical Institute (ETH) Z\"urich.  Weyl  included  his recent findings only  into the 3rd edition (1919) of the book. The English and French  versions \cite{Weyl:STM,Weyl:ETM}, translated from the  fourth revised edition (1921) contained  a short exposition of Weyl's generalized metric and the idea for a scale gauge theory of electromagnetism. 
E. Cartan read it and referred to it immediately.

Weyl's basic ideas for the generalization of  Riemannian metrics  in his papers of 1918 and in STM (3rd edition ff.) may be resumed as follows: 
\begin{itemize}

\item[(1)] Generalize Levi-Civita's concept of parallel displacement for Riemannian manifolds  to an abstract kind of ``parallel displacement'', not a priori linked to a metrical structure, $\Gamma = (\Gamma ^i_{j k})$,  called  {\em affine  connection}  (in { later} formed Cartanian terms: {\em  torsion free linear} connection).  

\item[(2)] Build up geometry from the { purely infinitesimal} point of view (``local'' in today's physicists language, i.e., using essentially the tangent structure of the manifold), with 
{\em similarities} as the basic transformations of space structure, because    {\em no natural unit} should be assumed in geometry a priori.
\item[(3)] The  possibility to   compare directly metrical quantities (physical observables) at different points of the spacetime  manifold $M$  ought to be considered a   {\em defect of Riemannian geometry} which is due to its historical origin in Gaussian surface theory. It presupposes a kind of ``distant geometry'' counter to modern field physics. 
\end{itemize}

In Weyl's view it should be possible to choose  a  {\em scale} (Ma\ss{}stab) freely and independently at every point of spacetime $M$; that  meant to  {\em gauge} the manifold. Then one arrives at a Riemannian (or Lorentzian etc.) metric
$ g:= ( g_{\mu \nu })$ with the squared line element  
\[ ds^2 = \sum g_{\mu \nu } dx^{\mu } dx^{\nu }.\]  
Let us call it   the  {\em Riemannian component} of a gauged {\em Weylian metric}. 
Comparison of quantities (observables) at different points was then possible only by  { integrating}  a {\em length} or  {\em scale connection}, given by a differential 1-form,
\[ \varphi = (\varphi_{\mu }) \quad  \quad \;\; \varphi = \sum \varphi_{\mu } dx^{\mu } = \varphi_i dx^{\mu } \, , \]
which expresses the infinitesimal change of measuring standards (relativ to the gauge). 
Both components together  ${(g, \varphi)}$ specify the metric in the chosen gauge.

To secure consistency, a different choice of the scale $\tilde{g} = \Omega^2 g$ has to be accompanied by a  transformation  
\beq  \tilde{\varphi} = \varphi - d (\log \Omega) = \varphi - \frac{d \Omega}{\Omega} \, ,
  \eeq
 a {\em gauge transformation (Eichtransformation)} in the literal sense of the word.  In late 1918 this word appeared in the correspondence with Einstein  \cite[VIII,  661]{Einstein:CPAE}, maybe after their oral discussion in the months before. In  1919 Weyl started to use it in his publications. 

In moderately modernizing language we may consider a {\em Weylian metric} $[(g, \varphi)]$ to be defined by an equivalence class of pairs  ${(g, \varphi)}$. Equivalence is given by gauge transformations. 

With this generalization of Riemannian geometry, Weyl looked for gauge covariant descriptions of  properties and in particular for  {\em gauge invariant}  objects  among which  the {\em scale curvature} (curvature of the scale connection)  $ f := d \varphi$ was the first to detect. He found that  a Weylian metric uniquely determines a compatible  affine   connection, the  (scale gauge dependent) Weyl-Levi-Civita connection $\Gamma = \Gamma (g,\varphi)$. It leads to   scale {\em invariant} 
 Riemann and Ricci curvatures  $Riem, \, Ric$
  and  scale {\em invariant} geodesics. 
A Weylian metric turned out to be reducible to a Riemannian one, iff $f= d\varphi=0$ ({integrable Weyl geometry}). 
 Finally, Weyl derived a tensor $C = (C_{ijkl})$ depending only on the conformal class $[g]$  of the metric, with  $C=0$ a necessary condition for conformal flatness (not  sufficient) if $dim\, M = n > 3 $. Later it was called {\em conformal curvature} or  {\em Weyl tensor} 
\cite[21]{Weyl:InfGeo}.
 
As already mentioned, Weyl originaly   identified the scale connection 
 $ { \varphi }$  with the  potential of  the {\em electromagnetic} (em) field. That led 
 to a  {\em gauge field} theory for electromagnetism  with group $(\R^+, \cdot)$. He thus  thought that 
the  Weylian metric $[(g, \varphi)]$ was able to  {\em unify } gravity and em  interaction. 
In this frame  the  Mie-Hilbert theory of matter with its combined Lagrangian  for gravity and electromagnetism could be placed in a geometrically unified scheme. This would, so Weyl hoped for roughly two years,  then lead to a success for a  purely field theoretic,  {\em dynamistic} theory of matter.

{ Einstein did not trust} Weyl's new theory physically, although he admired it from a mathematical point of view. He praised  the ``beautiful consequence
(wunderbare Geschlossenheit)'' of Weyl's thought ``\ldots  {\em apart from its agreement with reality} \ldots '' (emphasis, ES) \cite[vol. VIII, letter 499]{Einstein:CPAE}. For 
Einstein  the path dependence of  the scale transfer function for the measurement units 
\beq \lambda (p_0, p_1) = e^{\int_0^1 \varphi (\gamma ')} d\tau \quad \gamma \; \mbox{path from $p_0$ to $p_1 $ }  
\eeq 
gave reason to serious concern. In his view   no stable frequency of atomic clocks could be expected in Weyl's theory. But Weyl was not convinced. He countered by the  assumption that 
there seems to be a  {\em natural} gauge for atomic clocks because they adapt to the local field constellation of  scalar curvature (Weyl gauge).   

Other physicists, among them 
A. Sommerfeld, W. Pauli, and A. Eddington,  reacted differently and at first  positively. 
But after a period of reconsideration they also adopted a more critical position. That did not remain without influence on Weyl.
In particular Pauli's critique formulated in his article on general relativity in the {\em Enzyklop\"adie Mathematischer Wissenschaften}      \cite{Pauli:1921}, known to Weyl in draft already in summer 1920, and during discussions at Bad Nauheim in September the same year left traces on Weyl's position. 

In late 1920  { Weyl  withdrew} from defending his program of purely { field theoretical explanation of matter} and  {relativized} the  role of his  unified field theory. But he   did  { not}  give up his  program of {\em purely infinitesimal geometry}.

\subsection*{What remained?}
Weyl's ideas contained two germs of insight which turned out of long lasting importance:
\begin{itemize}
\item The enlargement of the automorphism group of classical differential  geometry by the scale gauge group resulted in   a {\em new invariance principle}.  Weyl identified it as ``the law of the conservation of electricity'' \cite[38]{Weyl:GuE}. 

\item Moreover, scale gauge geometry was conceptually basic and structurally well founded. Weyl showed this in an investigation which he called  the  {\em analysis of the problem of space (APOS)}
\end{itemize}
 
The first point was later identified as a special case of {\em E. Noether's theorems} \cite{Noether:1918}.\footnote{Noether's paper {\em Invariante Variationsprobleme} was presented the 26. 07. 1918  to the  G\"ottingen Academy of Science  by F. Klein; the final version appeared in  September 1918. Weyl could not know it in his publications \cite{Weyl:GuE,Weyl:InfGeo}. He referred to variational considerations by Hilbert, Lorentz, Einstein, Klein and himself. This remained so even in his later publications \cite{Rowe:Noether}}
With  Yang/Mills' and Utiyama's generalization, it became an important structural feature of   
non-abelian gauge theory in the second half of the century. 
With regard to the second point, 
Weyl took up motifs of the 19-th century discussion of the  { problem of space} in the sense of {\em Helmholtz -- Lie -- Klein}  and adapted the mode of questioning to the constellation of field theoretic geometry after the rise of GRT. That made Weyl's enterprise compatible to \'Elie Cartan's broader program of an infinitesimal implementation of the Kleinian viewpoint.

\subsection*{Analysis of the problem of space  (APOS)}
Between 1921 and 1923 Weyl looked for deeper conceptual foundations of his purely infinitesimal geometry  in a manifold $M$ (the ``extensive medium of the external world'') as an a-priori characterization of the ``possible nature of space''.
 In a clear allusion to Kant's distinction of different kinds of statements  a priori, Weyl distinguished an  ``analytic part'' and a ``synthetic'' part of his investigation. In the first step, Weyl analyzed what he considered the  necessary features of any meaningful transfer of congruence considerations to purely infinitesimal geometry. In the second step he enriched the properties of the resulting structure by postulates he considered basic for a coherent geometric theory.  

His basic idea was that a group of generalized ``rotations'',   a (connected) Lie subgroup $G \subset SL_n \R$,  had to be considered similarly to Kleinian geometry. In the new framework of purely infinitesimal geometry, the group could no longer assumed to operate on the manifold $M$ itself, but had only ``infinitesimal'' ranges of operation. In slightly modernized terminology,  $G$ operated on every tangent spaces of $M$  separately. \\[-0.5em]

 {\sc Conceptually necessary features}   ({``Analytic part'' of APOS}):
\begin{itemize}
\item At each point  $p \in M$   point {\em congruences} (``rotations'')  $G_p  \subset SL_n \R$ are given. They operate on the infinitesimal neighbourhood of the point (in $T_p M$). All $G_p$ are isomorphic to   {some} $G \subset SL_n \R$.
\item The $G_p$ differ by conjugations from point to point
\[ G_p = h_p^{-1} G h_p  \, ,\]
where  $h_p$  lies in the normalizer $\tilde{G}$ of $G$ and depends on the point $p$. Weyl called   $\tilde{G}$  the  ``{similarity group}'' of $G$. 
\end{itemize}
The  $G_p$ allowed to speak of point congruences (``rotations'') inside each infinitesimal neighbourhood $T_pM$ only. In order to allow for a ``metrical comparison'' between two neighbourhoods of $p$ and $p'$, even for infinitesimally close points  $p$ and $p'$, another gadget was necesssary.  Weyl argued that the most general conceptual possibility for such a comparison was given by  a linear connection.
\begin{itemize}
\item  In addition to the $G_p$ a { linear connection } $\Lambda = (\Lambda ^i_{jk})$ is given (in general with torsion in the later terminology of Cartan). 
 Weyl called $\Lambda$ an  {\em infinitesimal congruence transfer}, or even simply a  (generalized) {\em metrical connection}. 
 \end{itemize}
An infinitesimal congruent transfer need not be ``parallel''. Thus an {\em affine connection}  $\Gamma$ (without torsion)\footnote{Weyl continued to call  $\Gamma$ a ``parallel transfer'', in distinction to the ``metrical'' transfer.} continued to play role different from  a general metrical connection. 
Moreover, two connections $ \Lambda ^i_{jk}$ and $ \tilde{\Lambda} ^i_{jk}$ may characterize the  same infinitesimal congruence structure. This is the case,  if they differ   (point dependently) by ``infinitesimal rotations'' from the Lie group of $G$. In modernized language that meant: 
\beq  \Lambda \sim \tilde{\Lambda} \Longleftrightarrow \Lambda - \tilde{\Lambda} = A ,\quad A \; \mbox{ diff. form with values in} \; \mathfrak{g}= Lie\, G\, .  \label{equivalence of Lambdas}  \eeq 

Rotations in the infinitesimal neighbourhoods and metrical connection were, according to Weyl, minimal conditions necessary for talking about infinitesimal geometry in a (generalized) metrical sense. He did not yet consider these conditions  sufficient, but established two additional postulates.

{\sc Complementary conceptual features} (``synthetic part'' of APOS):
In order that an infinitesimal congruence structure in the sense of the analytic postulates may  characterize the ``nature of space'' Weyl postulated that the following conditions are satisfied.
\begin{itemize}
\item  {\em Principle of freedom}:  
In a (specified sense, not discussed here in detail) $G$ allows the ``widest conceivable range of possible congruence transfers'' in one point.
\end{itemize}
With this postulate Weyl wanted to establish an infinitesimal geometric analogue to Helmholtz postulate of free mobility in the classical analysis of space. Of course, it had to be formulated in a completely different way. Weyl argued that  the ``widest conceivable'' range of possibilites for congruence transfers has to be kept open by the geometric structure, in order not to put restrictions on the distribution and motion of matter. In place of free mobility of rigid bodies Weyl put the idea of a free distribution of matter. 

The widest possible range for congruence transfer given, Weyl demanded from the group $G$ that it took care of a certain coherence of the infinitesimal geometric structure. For him such a coherence condition was best expressed by the existence of a uniquely determined affine connection among all the metrical connections which could be generated from one of them by arbitrary infinitesimal rotations at every point (cf. equ. (\ref{equivalence of Lambdas})).
\begin{itemize}
\item {\em Principle of coherence}:
To each congruent transfer $ \Lambda = (\Lambda  ^i_{jk})$ exists {\em  exactly one equivalent  affine} connection. 
\end{itemize}

In his Barcelona lectures  \cite{Weyl:ARP1923} Weyl gave an interesting argument by analogy to the constitution of a ``a state'' in which a postulate of freedom (for citiziens, rather than for matter in general) is combined with a postulate of coherence. He expected from the constitution of  a liberal republic  that the free activity  of the citizens is restricted only by the demand that it does not contradict  the ``general well-being'' of the community (the ``state''). So Weyl saw a structural analogy between the constitution of  a liberal  state and the  ``nature of space'' and used it to motivate the choice of the postulates of the  ``synthetic'' part of his analysis of the space.
 
 After a translation of the geometrical postulates into conditions for the Lie algebra of the groups which are able to serve as ``rotations'' of an infinitesimal congruence geometry in the sense of the  APOS (analytical and syntheic part ) Weyl managed, in an involved case by case argumentation, to prove the following  \\[0em]
{\bf Theorem:}
The only groups satisfying the conditions for ``rotation'' groups in the APOS (analytic and synthetic part)   are the special orthogonal groups of any signature, $G \cong SO(p,q)$  with ``similarites'' $\tilde{G}\cong SO(p,q) \times \R^{+}$ .\\[-0.7em]

That was a  pleasing  result for Weyl's  generalization of Riemannian metrics. It indicated that the structure of  { Weyl geometry} was not just one among many more or less arbitrary generalizations of Riemannian geometry, but of  basic conceptual importance.\footnote{In this sense, the analysis of the problem of space may also be read as a belated answer to another of Einstein's objections to accepting Weyl geometry as a conceptual basis for gravitation theory: Why should not appear a ``Weyl II'' who proposes to make angle measurement dependent on local choice of units? \cite[VIII, 777]{Einstein:CPAE}} 
Note that, in modernized language, the  ``similarities'' $\tilde{G}$, i.e. the normalizer in $GL(n)$ of the ``congruences'' $G$  plays the role of the structure group, not the ``rotations'' themselves. Weyl rather implemented a (normal) extension  of the congruence group as the structure group of his generalized ``metrical'' infinitesimal geometry. That gave place for the gauge structure characteristic for his approach.  

Accordingly  in the 4-th edition of  STM Weyl proudly declared that  the analysis of the problem of space ought to be considered ``\ldots a good example of the essential analysis [Wesensanalyse] striven for by phenomenological philosophy (Husserl), an example that is typical for such cases where a non-immanent esence is dealt with.'' \cite{Weyl:STM},  translation from  \cite[157]{Ryckman:Relativity}.

\subsection*{Weyl on conformal and projective structure in 1921}
Shortly after having arrived at the main theorem of APOS, Weyl wrote a short paper on the ``placement of projective and conformal view'' in infinitesimal geometry 
\cite{Weyl:projektiv_konform}. It was 
triggered by  a paper of Schouten which he had to review for   F. Klein. 
In this paper Weyl investigated classes of affine connections with the same geodesics. These defined a  {\em projective structure} (``projektive Beschaffenheit'') on a differentiable manifold. Weyl derived an invariant of the projective path structure, the   
  {\em projective curvature} tensor $\Pi$  of $M$. Vanishing of   $\Pi$  was  a condition for the  manifold to be  projectively flat. In this case it is locally isomorphic to a linear projective space.
  
  In addition, Weyl found a highly interesting relationship between conformal, respectively  projective differential geometry and a Weylian metric.\\[0.5em]
{\bf Theorem:} If two Weylian manifolds $(M, [(g, \varphi)])$,  $(M', [(g', \varphi ')])$  have identical conformal curvature, $C = C'$, and identical projective curvature, $ \Pi= \Pi '$, they are locally isometric in the Weyl metric sense. \cite{Weyl:projektiv_konform}\\[-0.5em]

This theorem, so Weyl explained, seemed to be of deep physical import. The conformal structure was the mathematical expression for the causal structure in a general relativistic spacetime. Physically interpreted, the projective structure characterized the inertial fall of mass points, independent of parametrization, i.e., independent of conventions for measuring local time.  Thus Weyl's theorem showed that {\em causal and inertial structure of spacetime}  uniquely determine its  {\em Weylian} ---  not Riemannian ---  {\em metric}. This observation was taken up by  Ehlers/Pirani/Schild half a century later in their famous paper  {\em The geometry of free fall and light propagation} \cite{EPS}. It made the community of researchers in gravitation theory aware of the  fundamental character of  Weyl metric structures  for gravity.

\subsection*{Outlook on Weyl in the later 1920s} In the  following years (1923 -- 1925)
 Weyl started his large research program in the representation theory of Lie groups \cite{Hawkins:LieGroups}. After an intermezzo of intense studies in philosophy of the mathematical sciences in late 1925 and 1926 \cite{Weyl:PMN}, he 
 turned towards the new quantum mechanics. He published his book on {\em Group Theory and Quantum Mechanics} \cite{Weyl:GQM} 
and, little later, on   the general relativistic theory of Dirac equation with a $U(1)$ version of gauge idea. This idea had been proposed, in different contexts by E. Schr\"odinger, F. London, O. Klein and  V. Fock.\footnote{\cite{Vizgin:UFT,Goenner:UFT,Scholz:Fock/Weyl}.}
In the early 1920s he started a correspondence with E. Cartan, interrupted for some years but taken up again in the year 1930. In the later phase of the correspondence the two mathematicians tried to find out inhowfar they could agree on the basic principles of infinitesimal geometry in the area dominated by the ideas of general relativity. We come back to this point at the end of this paper.

\section{Cartan }
\subsection*{Towards an infinitesimal version of Kleinian spaces}
 In 1921/1922 Cartan studied  the new questions arising from the theory of general relativity (GRT) for differential geometry. At that time 
 he could already build upon a huge expertise in  the theory of {\em infinitesimal Lie groups} (now {\em Lie algebras}),\footnote{Here we shall switch between the historical and the present terminology without discrimination.}
 which he had collected over a period of roughly thirty years. Among others  he   had classified the simple complex Lie groups in \cite{Cartan:1894},  twenty years later the real ones \cite{Cartan:1914}. Moreover he had  brought to perfection the usage of    {\em differential forms} (``Pfaffian forms'')   in  differential geometry  \cite{Katz:DiffForms}. In 1910 he had started to describe the differential geometry of  classical motions by generalizing Darboux' method of
  ``tri\`edres mobiles'' (moving frames)  \cite{Cartan:1910}.\footnote{For a more detailed discussion of the following see \cite{Nabonnand:Cartan_2009}.}

In the early 1920s Cartan turned towards  reshaping the Kleinian program of geometry from an infinitesimal geometric point of view. 
In several notes in the {\em Comptes rendus} he first announced his ideas of how to use  {\em infinitesimal group structures}  for studying the foundations of GRT.
Different from Weyl and most other authors, he did  {\em not} rely  on the  ``absolute calculus'' of Ricci/Levi-Civita. He rather  built, as much as possible, on his calculus of differential forms. Starting from Levi-Civita's {\em parallel displacement} like Weyl,  he generalized  this idea  to {\em connections}  with respect to  various groups and 
 devised a general method for  differential geometry, which transferred Klein's ideas of the  Erlangen program to the infinitesimal neighbourhood  in  a differentiable manifold. These were  ``glued'' together by the generalized connection in such a (``deformed'') way, that the whole collection did not, in general, reduce to a classical Kleinian geometry.  The arising  structures were  later to be called  {\em Cartan geometries} \cite{Sharpe}.

\subsection*{Deforming Euclidean space} 
Before Cartan could ``deform'' Euclidean space $\E^3$, the latter had first to be {\em analyzed} in the literal sense of the word. That is, the homogeneous space $\E^3 \cong Isom\,\E^3/SO(3,\R)$  was thought to be disassembled into infinitesimal neighbourhoods  bound together by a connection, such that from an integral point of view classical Euclidean geometry was recovered. In a second step, the arising structure could be deformed to a more general infinitesimal geometry.
 
In order to analyze Euclidean space with  coordinates  $x= (x_1, x_2, x_3)$   Cartan  postulated that 
\begin{itemize}
\item orthogonal   {\em $3$-frames} (``tri\`edres'' -- triads) $( e_1(x), e_2(x), e_3(x) )$,    be  given at every  point $A$;
\item frames in an ``infinitesimally close point'' $A'$ (described in oldfashioned notation by coordinates  $x+dx$) may be related back to the one in $A$ by (classical) parallel transport. Cartan expressed that by differential 1-forms
\[  \omega _1, \omega _2, \omega _3, \; \;  \omega _{ij}=-\omega _{j i} \quad (1\leq i, j\leq 3) \, . \]
\end{itemize}
In total, $\omega = ( \omega _1,\omega _2,  \omega_3,    \omega _{12}, \omega _{13},\omega _{23})$ obtained values in the infinitesimal {\em inhomogeneous Euclidean group} $\R^3 \oplus so(3)$.\footnote{The infinitesimal displacement $dx =(dx_1, dx_2, dx_3)$ from $A$ to $A'$ is described by a tangent vector $\sum \omega^i e_i$. The $\omega^i$ are differential 1-forms dual to the $e_i$ (they depend linearly on the $dx_j$). The change of orthogonal frames in $A$ to frames $e'_1, e'_2,e'_3$ in $A'$ is described by an infinitesimal rotation, $e_i = \sum \omega_i^{\, j} e_j$  ($(\omega_i^{\, j})$ element of the Lie algebra $ \mathfrak{so}(3)$), the entries of which not only  depend linearly on $dx_k$ but also on the parameters of the rotation group (written by Cartan as $x_3, x_4, x_6$). \label{fn omegas} }

Cartan knew that in Euclidean space  the $\omega $s had to satisfy a compatibility condition
\[  \omega _i' = \sum_{k}\, [\omega _{k} \omega _{ki}] \, ; \quad   \omega _{ij}' = \sum_{k}\, [\omega _{ik} \omega _{kj} ] \, . \]
He called this the {\em structure equation} of Euclidean space (later {\em Maurer-Cartan)} equation.   
Here $\omega _i'$  denoted the exterior derivative of the differential form and square brackets the alternating product of differential forms. 

Using upper and lower index notation  $\omega ^i$ and $\omega _j^{\; k} $ for the differential forms 
 and Einstein's summation convention,    the equation may be rewritten as
\beq   d\omega ^i  =   \omega ^k \wedge  \omega _k^{\;i}    \, , \quad \quad 
        d\omega _i^{\; j}  =   \omega _i^{\; k} \wedge \omega _k^{\; j}  \, .  \label{Euclidean space}   \eeq 
        
        Passing to ``deformed Euclidean space'', Cartan allowed for the possibility that parallel   parallel transport of the triads around an  infinitesimal closed curve may result in an  
``infinitesimal small translation''  and/or an infinitesimal ``rotation''  \cite[593f.]{Cartan:1922[58]}. 
Then the  {\em structure equations} were generalized and became, denoted in moderately modernized symbolism,
\beqarr  d\omega ^i  &=&   \omega ^k \wedge  \omega _k^{\;i}  + \Omega ^i   \, \label{torsion} \\ 
      d\omega _i^{\; j}  &=&   \omega _i^{\; k} \wedge \omega _k^{\; j}  +  \Omega _i^j  \; ,\quad \quad  \ \label{curvature}
\eeqarr 
with  differential 2-forms $\Omega ^i  $  (values in the translation part of the Euclidean group)  and $\Omega _i^j $ (rotational part), which describe the deviation from Euclidean space. Cartan called them the   {\em torsion}  (\ref{torsion}) and  {\em curvature} form (\ref{curvature}) respectively.

\subsection*{Cartan spaces in general}
A little later,  Cartan went a step further and generalized his approach of deforming Euclidean spaces to other homogeneous spaces. The underlying idea was:
\begin{quote}
One notices that what one has done for the Euclidean group, the structural equations of which [(\ref{Euclidean space}) in our notation, E.S.] have been deformed into [(\ref{torsion}, \ref{curvature})], can be repeated for any finite [dimensional]  or infinite [dimensional]  
group.\footnote{``On con\c{c}oit que ce qui a \'et\'e fait poir le groupe euclidien, dont les \'equations de structure (1) sont d\'eform\'ees en (1'), peut se r\' ep\'eter poir n'importe quel groupe, fini ou infini.''}  
  \cite[627]{Cartan:1922[61]}
\end{quote}

As announced in this programmatic statement, Cartan 
 studied  diverse different ``spaces with connections'' or ``non-holonomous spaces''  (later terminology {\em Cartan spaces}) during the following years.\footnote{The terminology 
``non-holonomous'' was taken over from the specification of constraints in classical mechanics, see \cite{Nabonnand:Cartan_2009}. }  
Cartan's spaces $M$ arose from ``deforming'' a classical homogeneous space $S$ with Lie group $L$ acting transitively and with isotropy group $G$, such that
\[ S\approx L/G \, . \]
He directed his interest on the infinitesimal 
 neighbourhoods in $S$  described, in modernized symbolism, by
\[ \frak{l} / \frak{g} \cong  \frak{k} \, \quad \mbox{with}  \quad \frak{l}= \text{Lie}\, L, \; \frak{g} = \text{Lie}\, G \; ,\]
 $\frak{k}$ an infinitesimal sub-``group'' (i.e., subalgebra of $\frak{l}$),  invariant under the adjoint action of $G$.\footnote{Compare the modern presentation of Cartan geometry in \cite{Sharpe}.} 

The ``deformation'' of a Kleinian geometry in $S\approx L/G $  presupposed  identifications of a typical infinitesimal neighbourhood of $S$  with the infinitesimal neighbourhoods of any point of a manifold $M$ (Cartan:  ``continuum'') which was used to  parametrize the deformed space. Cartan thought about such identification in terms of smoothly gluing  homogeneous spaces $S$ to any point $p\in M$. More precisely,  
  $\frak{k}$ had to  be ``identified'' with $T_x M$ for all points $x \in M$ in such a manner  that the transition to an infinitesimally close  point $p'$  could be related to the $T_pM$  sufficiently smoothly. Such an identification was not always without  difficulties, although in general 
Cartan  presented  the transformation group $L$  as operating on  a (properly chosen) class of ``reference systems'' 
(``r\'ep\`eres'') and could derive such an identification from the infinitesimal elements in the  ``translational'' part of $L$.\footnote{See  the discussion with Weyl below.}  
These intricacies left aside here, a connection 1-form $\omega $ on $M$ with values in $\frak{l}$ could  be used to define a connection in the  infinitesimalized Kleinian geometry. Then the structural equations (\ref{torsion}), (\ref{curvature}) defined torsion and curvature of the  respective ``non-holonomous'' (Cartan) space. 

In particular, Cartan  studied non-holonomous spaces of the  
 \begin{itemize}
\item Poincar\'e group in papers on the geometrical foundation of  general relativity \cite{Cartan:1922[61],Cartan:1923[66],Cartan:1924[69]}  (for torsion  $\Omega ^i =0$ such a Cartan space reduced to a Lorentz manifold and could be used for treating Einstein's theory in Cartan geometric  terms),
\item  inhomogeneous similarity group (for torsion $ =0$, this case reduced to Weylian manifolds),
\item  conformal group \cite{Cartan:1922[60]},
\item  projective group \cite{Cartan:1924[70]}.
\end{itemize}

In the last case, Cartan introduced barycentric reference systems in infinitesimal neighbourhoods of a manifold (tangent spaces $T_pM$) (``r\'ep\`eres attach\'es aux differentes point de la vari\'et\'e'') and considered projective transformations of them. He remarked  that this is possible in `` \ldots  infinitely many different ways according to the choice of the reference systems''.\footnote{ ``\ldots une infinit\'e des mani\`eres different suivant le choix de r\'ep\`eres''. 
Translated into much later language, Cartan hinted here at the  possibility of different trivializations of the projective  tangent bundle.}
 That came down to considering the  projective closure of all tangent space.  
 
 In this way, Cartan developed an impressive  conceptual frame for studying different types of differential geometries, Riemannian, Lorentzian, Weylian, affine, conformal, projective, \ldots All of them were not only characterized by  connections and curvature, but enriched  by the possibility to allow for the new phenomenon of torsion. And all of them arose from Cartan's unified method of adapting the Kleinian viewpoint to infinitesimal geometry.

 \subsection*{Cartan's space problem }
Cartan learned to know about Weyl's problem of space  from the French translation of STM \cite{Weyl:ETM} and gave it his own twist 
\cite{Cartan:1922[62],Cartan:1923[65]}.
He  tried to make sense of Weyl's   descriptions of how the ``nature of space'' ought to be characterized by  ``rotations'' operating in infinitesimal neighbourhoods in terms of his own concepts.
He interpreted Weyl's vague description of the ``nature of space'' to mean a class of 
non-holonomous spaces with isotropy group $G \subset SL_n \R $ and the corresponding inhomogeneous group  $L\cong G \ltimes \R^n $. 

Cartan understood Weyl's ``metrical connection''in the sense of a class of  (Cartan) connections $[\omega ]$ with regard to $G,$ respectively $L$,  where two exemplars  of the class, $\omega , \, \bar{\omega } \in [\omega ]$, differed by a 1-form with values in $\mathfrak{g}$ only. 
 That was a plausible restatement of the ``analytical part'' of Weyl's discusssion; but   Cartan passed without notice over Weyl's  distinction between ``congruences'' ($G$) and ``similarities'' ($\tilde{G}$). So he  suppressed the specific group extension  (basically $\tilde{G} = G \times \R^+$), which led to Weyl's  scale gauge structure.

On that background Cartan reinterpreted Weyl's ``synthetic'' part of the analysis and stated
\begin{itemize}
\item   `` le premier axiome de  M. H. Weyl'': In any class $[\omega ]$  defining a (``metrical'') connection with values in $L$, one can find one connection with {\em torsion} $=0$. 
\item 
`` le second axiome de  M. H. Weyl'': Every class $[\omega ]$  gives rise to only one torsion free connection.
\end{itemize}

Cartan's rephrased ``premier axiome'' had, in fact, not much to do with Weyl's postulate of freedom, but at least it was an attempt to make mathematical sense of it. 
Using his knowledge in classification of infinitesimal Lie groups, he could argue that the ``first axiom'' is satisfied  not only by the generalized special orthogonal groups $SO(p,q)$ but also by the special linear group itself, the symplectic group (if $n$ is even), and by the largest subgroup of $SL_n \R$ with an invariant 1-dimensional subspace \cite[174]{Cartan:1923[65]}. If the second axiom was added, only the special orthogonal groups remained 
\cite[192]{Cartan:1923[65]}.

Cartan's simplification avoided the subtleties and vagueness of Weyl's ``postulate of freedom''.   Together with the streamlining of the analytical part of the analysis, he arrived at a slightly modified characterization of the problem of space. In this form it was transmitted to the next generation of differential geometers and entered the literature as {\em Cartan's problem of space} (S.S. Chern, H. Freudenthal, W. Klingenberg,  Kobayashi/Nomizu).

In the 1950/60s Cartan's space problem was translated into the fibre bundle language of modern differential geometry without the use of Cartan spaces. In these terms, 
 an  $n$-frame bundle over a differentiable manifold $M$, with group reducible to  
$G \subset SL_n \R$, was called a {\em  $G$-structure} on $M$. In $G$-structures   linear connections with and without torsion could be investigated.
The central question of the {\em Cartan-Weyl space problem} (i.e., the  Weylian space problem in Cartan's reduced form) turned into the following:  Which groups $G\subset SL_n \R$ have the property that every  $G$-structure carries  exactly one torsion free connection ?

It turned out that the answer was essentially the one given by Weyl and Cartan, i.e. the generalized special orthogonal groups of any signature, with  some additional other special cases \cite[vol. II]{Kobayashi/Nomizu}.
From the group theoretical point of view these considerations were still  closely related to Weyl's problem of space, while the geometrical question had now been modified twice, first by Cartan, then by the differential geometers of the next generation. 
  Only a minority of authors was still aware of the difference between Weyl's and Cartan's problem of space \cite{Scheibe:Spacetime,Laugwitz:ARP}. These authors insisted that it ought not to be neglected from a geometrial point of view .

 \subsection*{Toronto talk: Erlangen,  Riemann, and GRT }
At  the International Congress of  Mathematicians 1924 at Toronto, Cartan found an occasion to explain  his view of differential geometry in a clear and intuitive way to a broader mathematical audience. He started from a reference to the classical problem of space in the sense of the late 19th century:
\begin{quote}
From M. F. Klein (Erlangen program) and S. Lie one knows the important role of group theory in geometry.
 H. Poincar\'e popularized this fundamental idea among the wider scientific public \ldots \\[0.2em]
\ldots In each geometry one attributes the properties [of figures] to the corresponding group, or {\em fundamental} group [Hauptgruppe] \ldots
\end{quote}

  It was clear, however, that Riemann's ``M\'emoire c\'el\`ebre: Ueber die Hypothesen, welche der Geometrie zu Grunde liegen''  stood in stark contrast to such a perspective.
\begin{quote}
At first look, the notion of group seems alien to the geometry of Riemannian spaces, as they do not possess the homogeneity of any space with [Hauptgruppe]. In spite of this, even though a Riemannian space has no absolute homogeneity, it does, however, possess a kind of infinitesimal homogeneity; in the immediate neighbourhood it can be assimilated to a [Kleinian space]. \ldots
\end{quote}
 Such an  ``assimilation'', as understood by him, stood in close connection to frames of references or, in the language of physics, to observer systems in relativity.  Cartan observed:
\begin{quote}
[T]he theory of relativity faces the paradoxical task  of interpreting, in a non-homogeneous universe, all the  results  of so many experiences by observers who believe in homogeneity of the universe.  This  development has partially filled the gap which separated  Riemannian spaces from Euclidean space (``qui permit de combler en partie la fosse qu s\'eparait les espace de Riemann de l'espace euclidien'')\ldots . \cite{Cartan:1924[73]}  
\end{quote}

Thus he did not hide the important role of general relativity for posing the question of how to relate the homogeneous spaces of the classical problem of space to the inhomogeneous spaces of Riemann. But while in physics and philosophy of physics the debate on the changing role of ``rigid'' measuring rods or even ``rigid'' bodies was still going on, Cartan himself had been able  to ``fill the gap which separated  Riemannian spaces from Euclidean space'' in his own work ---  building upon the work of  Levi-Civita and his own expertise in Lie group theory and differential forms. That was similar to what Weyl had intended; but Cartan devised  a quite general method for constructing  finitely and globally inhomogeneous spaces from infinitesimally homogeneous ones. In the result, Cartan achieved a reconciliation of the Erlangen program and Riemann's differential geometry on an even higher level than Weyl had perceived.

\section{Discussion Cartan -- Weyl (1930)}
\subsection*{Weyl's Princeton talk 1929}
In June 1929 Weyl visited the United States and used the occasion to make Cartan's method  known among the Princeton group of differential geometers. O. Veblen and T.Y. Thomas had started to study projective differential geometry from the point of view of path structures \cite{Veblen:1928,Thomas:1926,Thomas:Trends}. To bring both viewpoints in connection, Weyl outlined Cartan's approach of infinitesimalized Kleinian geometries. He  discussed, in particular, how to identify  Cartan's generalized ``tangent plane'', the infinitesimal homogeneous space $\mathfrak{k}$  in the notation above, with the tangent spaces $T_pM$ (``infinitesimal neighbourhood'' of $p$) of the differentiable manifold $M$. For making the Princeton view comparable with Cartan's, one needed  not only that an isomorphism  $\mathfrak{k} \longrightarrow T_pM  $  be given for every point $p\in M$.   Weyl argued that one  even needed a contact condition of higher order (``semi-osculating'') \cite[211]{Weyl:1929[82]}. In this case a torsion free projective  connection, in the sense of Cartan, was  uniquely characterized by a projective path structure studied by the Princeton group (leaving another technical condition aside). 
 
\subsection*{Cartan's disagreement}
 Cartan was not content with   Weyl's presentation of his point of view. He protested in a letter to Weyl, written in early 1920:
\begin{quote}
Je prend connaissance de votre article recent (\ldots ) paru dan le Bulletin of the Amer. Math. Society. Je ne crois pas fond\'ee les critiques que vous addressez \`a ma th\'eorie des espace \'a connexion projective \ldots L'exposition que vous faites de ma th\'eorie ne r\'epond pas tout \`a faites \`a mon point de vue. \ldots  (Cartan to Weyl, 5.1.1930)
\end{quote}
 A correspondence of 3 letters between January and December 1930 followed.\footnote{The correspondence is preserved at ETH Z\"urich, Handschriftenabteilung,  \cite{Weyl/Cartan}.
 I thank P. Nabonnand for a giving me access to a transcription.}

 Cartan did not agree that an infinitesimal Kleinian space had to be linked as strictly to the tangent spaces $T_pM$ of the  manifold as Weyl had demanded. He defended a much more general point of 
view.\footnote{``En tous cas le probl\`eme d'\'etablir une correspondance ponctuelle entre l'espace \`a connexion projective et l'espace projectif tangent ne se pose ici pour moi: c'est un
 probl\`eme int\'eressant mais qui, dans ma th\'eorie, est hors de question'' \cite[Cartan to Weyl 5.1.1930]{Weyl/Cartan}.} 
He even went so far to admit a homogeneous space of different dimension from the base manifold.\footnote{``On pourrait m\^eme g\'en\'eraliser la g\'eom\'etrie diff\'erentielle projective \`a $n$ dimensions sur un continuum \`a $m \neq n $ dimension \ldots `` (Cartan to Weyl 5.1.1930). }
Thus Cartan tended toward what later would become fibre bundles over the manifold, here a projective bundle with fibres of dimension $n$ over a manifold of dimension $m$. On the other hand,  he also had studied the conditions under which the integral curves of  second order  differential equations  could be considered as geodesics  of a (``normal'') projective connection \cite[28ff.]{Cartan:1924[70]}

 Weyl  insisted even more on the {necessity} of a (``semi-osculating'') identification of the infinitesimal homogeneous space with the tangent spaces of the manifold, in order to get a differential geometric structure which would be truely {\em intrinsic}  to $M$. He reminded his correspondent that they had discussed this  question already in 1927, after a talk of E. Cartan at Bern:\footnote{\cite{Cartan:1927Bern}} 
\begin{quote}
I remember that we discussed this question alread at Bern, and that I was unable to make my point of view understood by you.  (Weyl to Cartan 24.11.1930)
\end{quote}

 In particular for the { conformal} and { projective structures} Weyl now saw great advantages of the studies of the Princeton group (Veblen, Eisenhart, Thomas). Apparently he came to the conclusion that they could be connected to  the Cartan approach only after such a smooth (semi-osculating) identification. 

Although he did not mention it in the discussion, it seems quite likely that  the  {physical import} of conformal (causal) and projective (inertial) structures for GRT  played an important  background  role for Weyl's insistence on the ``intrinsic'' study of conformal and projective structures.   In  1922 Weyl had realized that inertial/projective and causal/conformal structure together determine a Weylian metric uniquely (cf. end of section 2). 
Such considerations make sense, of course, only if conformal and projective structures are  understood  as intrinsic to the manifold.
 
\pagebreak
\subsection*{Trying to find a compromise}
 Although  Cartan at first defended  his more abstract point of view, he agreed that he might better have chosen a different terminology avoiding the intuitive language of a ``projective tangent space'', which he applied even in the more abstract case of fibre dimension different from $dim\, M$. 

After  Weyl had explained  why he insisted on the closer identification, Cartan became more  reconciliatory:
\begin{quote}
\ldots je vous accorde tr\`es volontiers. \ldots C'est un probl\`eme important et naturel de chercher comment l'espace lin\'eaire tangent est 'eingebettet' dans l'espace non-holonome donn\'e. (Cartan to Weyl, 19.12.1930)
\end{quote}

 At the end of the year,  after the initial problems to understand each other had been resolved,   Cartan  admitted    that Weyl's question was not just any kind of specification inside his more general approach. Cartan's general view was neither withdrawn nor  devaluated; it later found its extension in the theory of fibre bundles.  But for the more intrinsic questions of differential geometry the identification of infinitesimal Kleinian geometry with the tangent space of the base manifold has become part of the standard definition of {\em Cartan geometry}. 

\section{In place of a resum\'ee}
Weyl and Cartan started from quite different vantage points for the study of generalized differential geometric structures motivated by the rise of general relativity.  Both put infinitesimal group structures in the center of their considerations. In the early 1920s Cartan had a lead over Weyl in this regard, and it were exactly such geometrical considerations which led Weyl into his own research program in Lie group representations \cite{Hawkins:LieGroups}. After he came into contact with Einstein's theory, Cartan immediately started to work out a general framework how differential geometry could be linked to an infinitesimalized generalization of Klein's Erlangen program. 

Weyl, on the other hand, started from a natural philosophically motivated generalization of Riemannian geometry which, as he hoped for about two years, might  be helpful for  unifying gravity and electromagnetism and might help to solve the riddle of a field theoretic understanding of basic matter structures. After he began to doubt the feasibility of such an approach, he turned towards a more general conceptual-philosphical underpinning of his geometry. That led him to take up the analysis of the problem of space from the point of view of infinitesimal geometry. 

Both authors agreed upon the importance of using infinitesimal group structures for a generalization of differential geometry in the early 1920s. They read the work of each other and managed to come to grips with it, even though sometimes with difficulties and with certain breaks. Still at the end of the 1930s Weyl admitted, in an otherwise very positive and detailed review of Cartan's recent book \cite{Cartan:1937}, the problems he had with reading Cartan.\footnote{``Does the reason lie only in the great French geometric tradition on which Cartan draws, and the style and contents of which he takes more or less for granted as a common ground for all geometers, while we, born and educated in other countries, do not share it?'' \cite[595]{Weyl:Review_Cartan} }
But in spite of differences with regard to technical tools and emphasis of research guidelines, they came to basically agree on the way, how connections in various groups could be implemented as basic conceptual structural tools in  the rising ``modern'' differential geometry of the second third of the new century.

\nocite{Deppert:Kiel}
\small
\bibliographystyle{apsr}
\bibliography{a_litfile}



\end{document}